# Probing single protein dynamics on liposome surfaces

Dong-Fei Ma[1,4], Chun-Hua Xu[1], Wen-Qing Hou[1,4], Chun-Yu Zhao[2,4], Lu Ma[1], Cong Liu[2], Jiajie Diao[3], Ying Lu[1*] and Ming Li[1,4*]

[1] Beijing National Laboratory for Condensed Matter Physics and CAS Key Laboratory of Soft Matter Physics, Institute of Physics, Chinese Academy of Sciences, Beijing 100190, China.

[2] Interdisciplinary Research Center on Biology and Chemistry, Shanghai Institute of Organic Chemistry, Chinese Academy of Sciences, Shanghai 200032, China.

[3] Department of Cancer Biology, University of Cincinnati School of Medicine, Cincinnati, OH 45267, USA.

[4] University of Chinese Academy of Sciences, Beijing 100049, China.

*Supporting Information Placeholder*

**ABSTRACT:** It is crucial to measure position and conformational changes of a membrane-interacting protein relative to the membrane surface. This is however challenging because the thickness of a membrane is usually only about 4 nm. We developed a fluorescence method which makes use of the principle of FRET between a fluorophore and a cloud of quenchers encapsulated in a liposome, hence the name LipoFRET. LipoFRET can readily locate a fluorophore in different depths inside and at different heights above the membrane. We applied LipoFRET to study α-synuclein, a key player in the pathology of Parkinson's disease. Our approach yielded quantitative information about the dynamics of different regions of α-syn in lipid membranes, which has never been explored before.

Liposomes are widely used as model systems for studying interactions between lipids and proteins, and elucidating mechanisms of actions of drugs and antibiotics on target cells[1-2]. Applications of time- and ensemble-averaging techniques such as nuclear magnetic resonance (NMR) and electron paramagnetic resonance (EPR) to these model systems have provided many valuable data[3-4]. However, it is still challenging to gain information about position changes and structural dynamics of a membrane-interacting protein relative to the membrane surface. In general, a site of interest of a protein can be either embedded in the membrane or exposed to the aqueous solution around the liposome. There is however a lack of suitable techniques to detect the different positions. Assays employing environment-sensitive dyes can only report if a protein is inside or outside the lipid bilayer of the liposome[5-6]. FRET can be used to probe nano-scale movements of fluorophore-labeled proteins on liposomes[7-9], but it is useful only when the positions of quenchers are precisely known, which is not always readily to use because of the fluidity of the liposome membrane. Fluorescence-quenching by brominated lipids can measure the position of the protein relative to membrane surface when different brominated lipids are used in a series of measurements. It is not able to distinguish whether the fluorophore is in the inner or outer leaflet of the lipid bilayer[10-11]. Here we develop a liposome-based single-molecule method to resolve the challenge, and apply the method to characterize membrane binding of α-synuclein (α-syn), a key player in pathology of Parkinson's disease and presynaptic vesicle homeostasis.

The underlying physics of our method is the FRET between a single fluorophore and a multitude of quenchers encapsulated in a liposome (Figure 1a), hence the name LipoFRET. The transfer kinetics $k_t$ is the sum of pairwise transfer rates $k_{ti}$ (Figure S1 in the supporting information (SI))[12-14],

$$k_t = \sum_{i=1}^{N} k_{ti} = \frac{1}{\tau}\sum_{i=1}^{N}(\frac{r_{0i}}{r_i})^6, \qquad (1)$$

where $\tau$ refers to the lifetime of the donor, and $r_{0i}$ and $r_i$ are the Förster distance and the spatial distance between the $i^{th}$ donor-acceptor pair, respectively. The energy transfer efficiency is given by $E=k_t/(\tau^{-1}+k_t)$, which is used to calculate the relative fluorescence of the donor, $F/F_0=1-E$, where $F_0$ is the intrinsic fluorescence without the quenchers. Many dyes, and even metal ions, can be used as quenchers so long as their absorption spectra have some overlap with the emission spectrum of the donor. We chose trypan blue (TB)[15-16] and $Cu^{2+}$-nitrilotriacetic acid complex (Cu-NTA)[17] as quenchers in the current work.

Our calculations (Figure 1b) indicate that, to a good estimate, a distance change by ~1 nm will result in an intensity change by ~10% of $F_0$, which is associated with the Förster distance and the concentration of the quenchers. This is precise enough to measure position changes of a fluorophore-labeled protein near the lipid bilayer. Taking $r_0$=6.1 nm, which is the Förster distance between TB and Alexa555 according to spectra overlap calculations in SI, the region of sensitivity shifts from around the inner bilayer to around the out surface of the bilayer as the concentration of TB increases from 2.5 to 10 mM. The shifting is of importance for practical applications because the region of sensitivity of LipoFRET is usually a few nanometers only. If necessary, one may add some quenchers with short Förster distances[17] to attenuate the fluorophore only near the inner surface to enhance the sensitivity there (red and dark yellow lines in Figure 1b).

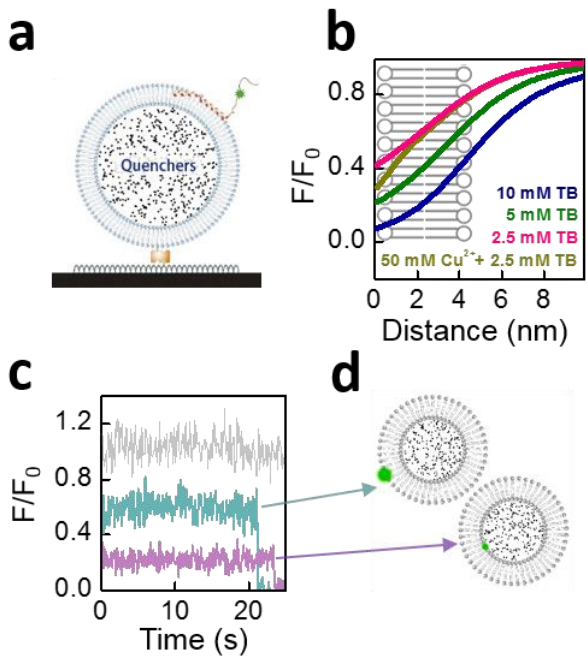

*Figure 1*. **Principle of LipoFRET.** (**a**) Schematic of LipoFRET to detect the position of a fluorophore around a liposome. (**b**) Quenching abilities vs. distance calculated for various quencher concentrations. (**c**) Fluorescence traces of fluorophores on liposomes' outer surface (green line) and inner surface (purple line) as depicted in (**d**). A trace for liposome without quenchers (gray lines) is also displayed for comparison.

We first demonstrated the feasibility of LipoFRET by measuring the quenching efficiency of fluorophores conjugated to lipid headgroups. We prepared unilamellar liposomes with the extrusion method[18] and immobilized the liposomes onto streptavidin-modified glass surface. The liposomes were doped with lissamine rhodamine B-labeled PE (Rhod-PE) with 0.001% molar fraction. For liposomes containing 2.5 mM TB, the fluorescence of Rhod-PE is significantly attenuated as predicted by our calculations (Figure S1d in SI, green line). Two intensities were observed, representing the fluorophores on the inner and outer resurfaces, respectively (traces in Figure 1c and scheme in Figure 1d). Therefore, LipoFRET is precise enough to measure the location of a single fluorophore in the lipid bilayer.

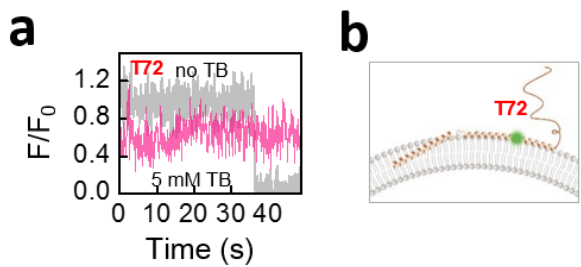

*Figure 2*. **Position of the residue T72 of α-syn.** (**a**) Typical fluorescent traces of α-syn labeled at T72 (sketched in (**b**)) on a liposome with (pink line) and without quenchers (gray line).

We choose α-syn as an example to show that LipoFRET is able to measure accurately the position of a protein relative to the membrane surface. To this end, we prepared unilamellar liposomes composed of DOPC/DOPA (molar ratio of 7:3), 16:0 biotinyl-cap PE (0.1% molar fraction), and used Alexa555-labeled α-syn in PBS buffer (pH=7.4, 150 mM NaCl) as the donors. The protein concentration was carefully controlled so that there was no more than one α-syn molecule on a liposome in most cases. As representatives, we selected three sites in α-syn to depict the general pattern of the protein on the liposome. They are S129 that locates on the C-terminal flexible acidic tail, T72 in the central alpha-helix, and K10 at the N-terminus. In each measurement, we first measured the intrinsic fluorescence $F_0$ of Alexa555-labeled α-syn on liposomes without the quenchers. The relative intensity of T72C-Alexa555 is $F/F_0$=0.64±0.12 (mean±s.d.) (Figure 2a and Figure S2 in SI) while that of S129C-Alexa555 is 0.87±0.13 (Figure. S2d and S2e in SI). Taking the thickness of lipid bilayer to be 4.5 nm,[19] the intensity of T72C-Alexa555 is consistent with our calculations (Figure 1b) assuming that α-syn adopts a helical conformation with T72 being on the hydrophilic side of the alpha-helix[4] (Figure 2b). For the site S129, the high fluorescent intensity in Figure S3 in SI indicates that this site is about 2.5 nm higher than T72 above the liposome surface. To the best of our knowledge, the distance between the membrane surface and any residues in the C- tail of α-syn has not been measured before, although it has long been proposed that the negatively charged acidic tail (residue 96-140) flaps in the aqueous milieu when its N-terminus anchors on a negatively charged lipid bilayer[3,20]. The homogenous aqueous environment around the unstructured tail poses an obstacle to almost any environment-dependent methods to determine its position above the membrane. LipoFRET is not subject to this limitation, therefore is suitable for detailed characterization at the single-molecule level in this case.

LipoFRET is capable of monitoring the position changes of a single protein when it interacts with a liposome. The interaction of the N-terminus of α-syn with lipid bilayer has been intensively investigated[20-22]. Some researchers reported that this region is fully buried in the acyl chains by using neutron reflectivity and bromine quenching[20-21], but others tended to believe that the whole helix stays only in the surface near the phospholipid headgroups[22]. In our measurements with liposomes containing 2.5 mM TB plus 50 mM Cu-NTA, the fluorescence of K10C-Alexa555 transits slowly among three values in a time scale of a few seconds (Figure 3, Figure S4). The mean dwell times of the N-terminus at the three states are almost equal (Figure 3c). The highest relative intensity $F/F_0$=0.84±0.11 (mean±s.d.) is similar to that of T72C-Alexa555. It is therefore located on the surface of the lipid bilayer. The one with medium intensity, $F/F_0$=0.68±0.11, is about 1.5 nm lower than T72. The lowest intensity is $F/F_0$=0.44±0.12, suggesting that the position is about 3.4 nm beneath the bilayer surface. To the best of our knowledge, this is the first quantitative data set about the dynamics of the N-terminus of α-syn in lipid membranes. Time- and ensemble-averaging techniques such as neutron reflectivity and tryptophan fluorescence quenching by bromine reported that α-syn penetrates the membrane by 1-1.5 nm[20-21]. We believe that those values were just the average of the three depths observed here.

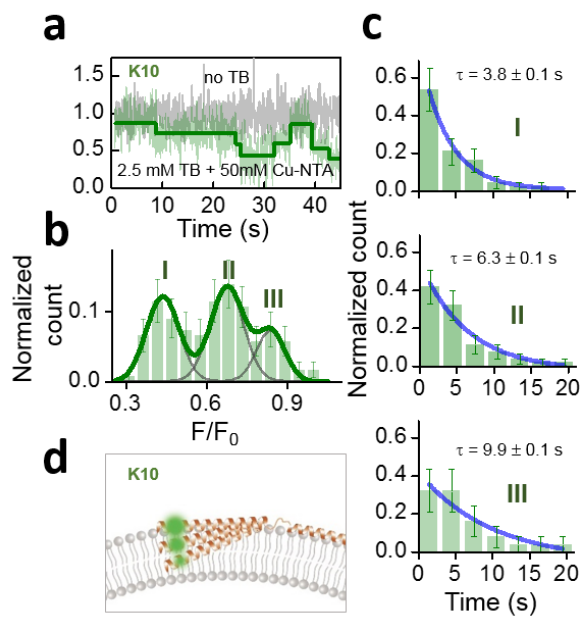
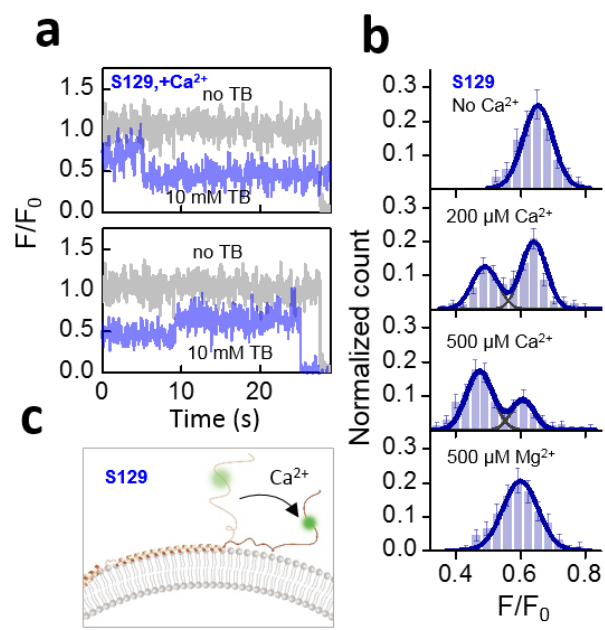

*Figure 3.* **The residue K10 of α-syn transits among three penetration depths.** (**a**) Typical traces. (**b**) Histograms of the fluorescent intensities. (**c**) Histograms of the dwell times of three positions. (**d**) Scheme of α-syn labeled at K10 on the liposome. Error bars are standard error in the bins. The statistics are from over 50 traces.

*Figure 4.* **Calcium regulation of the α-syn C-terminal tail on liposomes**. (**a**) Typical traces. (**b**) The intensity histograms of S129-Alexa555 in the presence of various concentrations of $Ca^{2+}$ (0 to 500 μM) are compared with that in the presence of 500 μM $Mg^{2+}$. (**c**) scheme of α-syn labeled at S129 without or with $Ca^{2+}$. The statistics are from over 150 traces at each concentration of $Ca^{2+}$.

Membrane proteins are usually not completely embedded in lipid bilayers because they have to interact with the surrounding aqueous environment to fulfil their functions[23]. So far, there is no such technique for detecting the position changes of site of interest in the protein located outside the membrane. Here we show that LipoFRET is able to investigate the position changes of the C-terminal acidic tail of α-syn which is exposed in the salt solution. The tail of α-syn is likely involved in the $Ca^{2+}$ binding process since α-syn is implicated functionally in dopamine and $Ca^{2+}$ signaling[24-26]. It was proposed that $Ca^{2+}$ can specifically interact with the C-terminal tail of α-syn, turning the tail from a solution-exposed state into a membrane-bound state[27]. We examined the effect of $Ca^{2+}$ on the acidic tail of the liposome-bound α-syn in 10 mM HEPES buffer (pH=7.4, 10 mM NaCl). When $Ca^{2+}$ was added, two fluorescent intensities were observed for S129C-Alexa555 (Figure 4), which correspond to two states with and without a $Ca^{2+}$ bound, respectively. The population of low fluorescence molecules are smaller than that of high fluorescence molecules in the experiment with 200 μM $Ca^{2+}$. The ratio is reversed at 500 μM $Ca^{2+}$. These results are consistent with previous observation that the half-saturation concentration of $Ca^{2+}$ binding to α-syn is about 300 μM[24]. In a control experiment with 500 μM $Mg^{2+}$, only one intensity is observable, which is almost equal to the high-fluorescent one in the experiments with $Ca^{2+}$ (Figure 4b). The shift to lower intensity of the high-fluorescence peak in Figure 3b results from the non-specific Coulomb screening of the C-terminal tail by $Ca^{2+}$. The same screening is also manifested in the control experiment with $Mg^{2+}$. Because of the Coulomb screening effect, the difference in height between the two states is 1.5±0.9 (mean±s.d.) nm at 200 μM $Ca^{2+}$ and 1.1±0.9 nm at 500 μM $Ca^{2+}$. Since $Ca^{2+}$ ions reduce the height of S129 to a much smaller value (Figure 4c), we speculate a specific interaction between the C-terminal tail and $Ca^{2+}$, resulting in a new conformation of the tail. An in-depth investigation is in progress to sketch the contour of the tail by measuring the positions of every site in the tail.

Probing the structural dynamics of proteins in lipid membranes had ever been a very difficult task. By using α-syn as an example, we demonstrated the feasibility of LipoFRET in quantitating positional changes of a site of interest in a protein not only inside but also outside the lipid membrane. Both locations are crucial to understanding protein-membrane and protein-ligand interactions. The results indicated that the spatial resolution of LipoFRET is ~1 nm. Previous studies showed that α-syn can partially bind to membranes, and different factors may influence the binding[28]. By performing LipoFRET, we revealed that α-syn displays an inclined conformation on the liposome surface, with its N-terminus being embedded in the bilayer and its C-terminus floating in the aqueous solution. Furthermore, our data showed that the N-terminus of α-syn transits among three positions spontaneously. The dwelling time of α-syn in each position is of the order of seconds, which is very slow as compared to the diffusive motion of the protein in the membrane. The existence of multiple states of α-syn in the membrane may explain why different results were reported in the literature. LipoFRET is easy to implement and does not require complicated instrumentation. Extending LipoFRET to other applications is straight forward. We anticipate more widespread applications of LipoFRET in the study of membrane systems.

## ASSOCIATED CONTENT

### Supporting Information

Experimental procedures, preparation of LipoFRET assay, protein preparation and labeling, calculation and measurement of system properties are included in the supporting information.
The Supporting Information is available free of charge on the ACS Publications website.

## AUTHOR INFORMATION

### Corresponding Author

Correspondence should be addressed to YL (yinglu@iphy.ac.cn) or ML (mingli@iphy.ac.cn).

## Author Contributions

ML and YL designed the project; DFM, CHX and WQH performed the experiments; SNZ and CL prepared the proteins; DFM, CHX, CL, JJD, YL and ML analyzed the data; DFM, JJD, CL, YL and ML wrote the manuscript.

## Notes

The authors declare no competing financial interests.

## ACKNOWLEDGMENT

This work was supported by a National Science Foundation of China (Grants No. 11674382 (to Y. L.), No. 91753104 (to M. L.) and No. 11574381 (to C. H. X)) and by the CAS Key Research Program of Frontier Sciences (Grant No. QYZDJ-SSW-SYS014 (to M. L.)). Y. L. is supported by the Youth Innovation Promotion Association of CAS (No.2017015).